\documentclass[a4paper,12pt]{ETHpaper}

\usepackage{graphicx}
\usepackage[dvipsnames]{xcolor}
\usepackage{dcolumn}
\usepackage{bm}
\usepackage{amssymb,amsmath}
\usepackage[british]{babel}
\usepackage{subfigure}
\usepackage{amssymb,amsmath,amsthm}
\usepackage[capitalise]{cleveref}
\usepackage[sort&compress,square,numbers]{natbib}
\usepackage{multirow}

\newcommand{\G}{\mathcal{G}}

\newcommand{\abs}[1]{\left| #1 \right|}

\newcommand{\AIC}{\text{AIC}}
\newcommand{\BIC}{\text{BIC}}

\begin{document}

\title{Analytical Formulation of the\\ Block-Constrained Configuration Model}
\titlealternative{Analytical Formulation of the Block-Constrained Configuration Model}
\author{Giona Casiraghi}
\authoralternative{G. Casiraghi}
\address{Chair of Systems Design, ETH Zurich, \\
Weinbergstrasse 56/58, CH-8092 Zurich, Switzerland \\
e-mail: gcasiraghi@ethz.ch}
\www{\url{http://www.sg.ethz.ch}}
\maketitle

\begin{abstract}
We provide a novel family of generative block-models for random graphs that naturally incorporates degree distributions: the block-constrained configuration model.
Block-constrained configuration models build on the generalised hypergeometric ensemble of random graphs and extend the well-known configuration model by enforcing block-constraints on the edge generation process.
The resulting models are analytically tractable and practical to fit even to large networks.
These models provide a new, flexible tool for the study of community structure and for network science in general, where modelling networks with heterogeneous degree distributions is of central importance.
	\paragraph{Keywords:} block model, community structure, random graphs, configuration model, network analysis, gHypEG 
\end{abstract}

\section{Introduction}\label{sec:intro}

Stochastic block-models (SBMs) are random models for graphs characterised by group, communities or block structures. 
They are a generalisation of the classical $G(n,p)$ Erd\"os-Renyi model~\cite{erdds1959random}, where vertices are separated into $B$ different blocks, and different probabilities to create edges are then assigned to each block.
This way, higher probabilities correspond to more densely connected groups of vertices, capturing the structure of clustered graphs~\cite{Fienberg1985,Holland1983,Peixoto2012a}.

SBMs are specified by defining a $B\times B$ block-matrix of probabilities $\mathbf B$ such that each of its elements $\omega_{b_ib_j}$ is the probability of observing an edge between vertices $i$ and $j$, where $b_i$ denotes the block to which vertex $i$ belongs.
Most commonly, block-matrices are used to encode community structures.
This is achieved by defining a diagonal block-matrix, with the inclusion of small off-diagonal elements.
Block-matrices, though, allow to define SBMs with a broad range of structures.
For example, by appropriately encoding the block-matrix $\mathbf B$ it is also possible to model core-periphery, hierarchical, as well as multipartite structures.

Thanks to its simple formulation, the edge generating process of standard SBMs is able to retain the block structure of the graph that needs to be modelled~\cite{Karrer2011}.
However, it fails to reproduce empirical degree sequences.
The reason for this is that in the $G(n,p)$ model and in its extensions, edges are sampled independently from each other, generating homogeneous degree-sequences across blocks.
This issue impairs the applicability of the standard SBM to most real-world graphs. 
Because of the lack of control on the degree distributions generated by the model, SBMs are not able to reproduce the complex structures of empirical graphs, resulting in poorly fitted models.

Different strategies have been formulated to overcome this issue.
Among others, one approach is that of using exponential random graph models~\cite{Krivitsky2012}.
These models are very flexible in terms of the kind of patterns they can incorporate.
However, as soon as their complexity increases they loose the analytical tractability that characterises the standard SBM.
Another practical approach taken to address the issue of uniform degree-sequences in SBMs are degree-corrected block models (e.g.~\cite{Peixoto2014a,Newman2015,Karrer2011,Peixoto2015x}).
Degree-corrected block-models address this problem extending standard SBMs with degree corrections, which serve the purpose of enforcing a given expected degree-sequence within the block structures.
On the one hand, the main advantage of degree-corrected block models is that they retain the simplicity of the standard SBM.
On the other hand, they usually lose the ability to generate graphs, as the degree correction is added only as a correction in the probability estimation of the model.

In this article, we propose a new family of block-models by taking a different approach.
By redefining the base edge generating process such that it preserves the degree sequence of the modelled graph we can avoid the need for degree corrections.
 We start from the simplest random model that can reproduce heterogeneous degree distributions: the configuration model of random graphs~\cite{Chung2002a,Chung2002,Bender1978,Molloy1995}.
The configuration model assumes that the number of edges in the graph is fixed, and it randomly rewires edges between vertices preserving the degree-sequence of the original graph.
We extend the standard configuration model to reproduce arbitrary block structures, by introducing block constraints on its rewiring process.
We refer to the resulting model as \emph{block-constrained configuration model} (BCCM).
The major advantages of our approach are (i) the natural degree-correction provided by BCCMs, and (ii) the fact that the model is still generative and analytically tractable.

\section{Generalised Hypergeometric Ensembles\\of Random Graphs (gHypEG)}\label{sec:ghype}

Our approach builds on the generalised hypergeometric ensemble of random graphs (gHypEG)~\cite{Casiraghi2016,Casiraghi2018}.
This class of models extends the configuration model (CM)~\cite{Molloy1995, molloy_reed_1998} by encoding complex topological patterns, while at the same time preserving degree distributions.
Block constraints fall into the larger class of patterns that can be encoded by means of gHypEGs.
For this reason, before introducing the formulation of the block-constrained configuration model, we provide a brief overview over gHypEGs.
More details, together with a more formal presentation are given in~\cite{Casiraghi2016,Casiraghi2018}.

\begin{figure}[h]
\centering
\includegraphics[width=.45\textwidth]{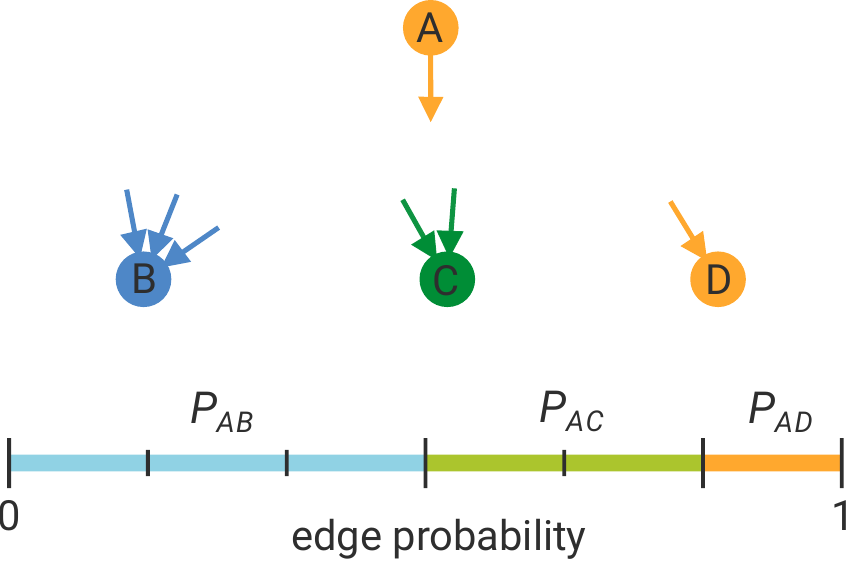}
\hfill
\includegraphics[width=.45\textwidth]{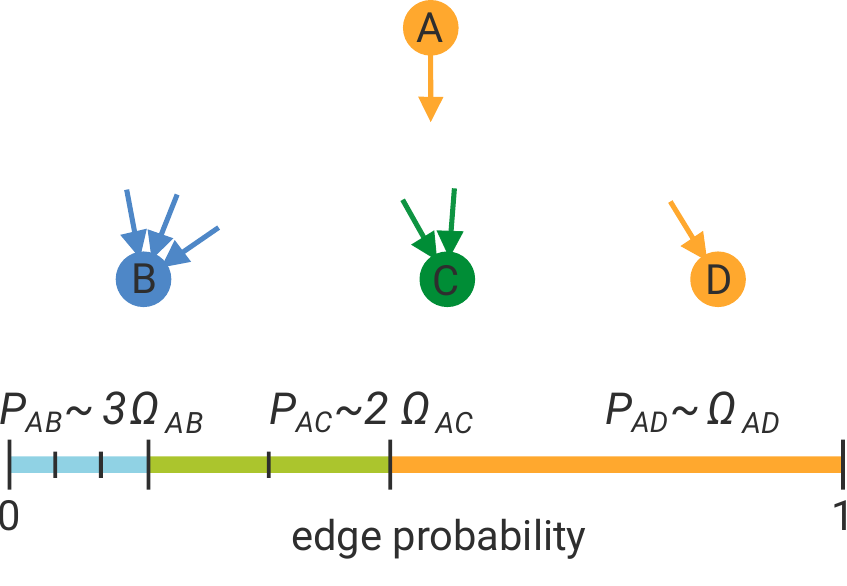}\hfill
\caption{\small Graphical illustration of the probability of connecting two vertices as a function of degrees (left figure), and degree and propensities (right figure).}
\label{fig:props}
\end{figure}

In the configuration model of random graphs, the probability to connect two vertices depends only on their (out- and in-) degrees.
In its most common formulation, the configuration model for directed graphs assigns to each vertex as many half-edges or out-stubs as its out-degree and as many half-edges or in-stubs as its in-degree.
It then proceeds connecting random pairs of vertices joining out- and in-stubs.
This is done by sampling uniformly at random one out- and one in-stub from the pool of all out- and in-stubs respectively and then connecting them, until all stubs are connected.
The left side of \cref{fig:props} illustrates the case from the perspective of a vertex $A$.
The probability of connecting vertex $A$ with one of the vertices $B$, $C$ or $D$ depends only on the abundance of stubs, and hence on the in-degree of the vertices themselves.
The higher the in-degree, the higher the number of in-stubs of the vertex, hence the higher the probability to randomly sample a stub belonging to the vertex.

Generalised hypergeometric ensembles of random graphs provide a closed form expression for the probability distribution underlying this process, where the degrees of the vertices are preserved in expectations.
This result is achieved by mapping the process described above to an urn problem.
Edges are represented by balls in an urn, and sampling from the configuration model is described by sampling balls (i.e., edges) from an urn appropriately constructed.
For each pair of vertices $(i,j)$, we can denote with $k^{\text{out}}_i$ and $k^{\text{in}}_j$ their respective out- and in-degrees.
The number of combination of out-stubs of $i$ with in-stubs of $j$ which could be connected to create an edge is then given by $k^{\text{out}}_ik^{\text{in}}_j$.
To map this process to an urn, for each dyad $(i,j)$ we should place exactly $k^{\text{out}}_ik^{\text{in}}_j$ balls of a given colour in the urn~\cite{Casiraghi2018}.
The process of sampling edges from the configuration model is hence described by sampling balls from this urn, and the probability distribution underlying the model is given by the multivariate hypergeometric distribution with parameters $\mathbf\Xi=\{k^{\text{out}}_ik^{\text{in}}_j\}_{i,j}$.

Generalised hypergeometric ensembles of random graphs further extend this formulation.
In gHypEGs, the probability to connect two vertices depends not only on the degree (i.e., number of stubs) of the two vertices, but also on an independent propensity of the two vertices to be connected, which captures non-degree related effects.
Doing so allows to constrain the configuration model such that given edges are more likely than others, independently of the degrees of the respective vertices.
This case is illustrated by the right side of \cref{fig:props}, where $A$ is most likely to connect with vertex $D$, belonging to the same group, even though $D$ has only one available stub.

In generalised hypergeometric ensembles the distribution over multi-graphs (denoted $\G$) is formulated such that it depends on two sets of parameters: the combinatorial matrix $\mathbf\Xi$, and a propensity matrix $\mathbf\Omega$ that captures the propensity each pair of vertices to be connected.
Each of these two matrices has dimensions $n\times n$ where $n$ is the number of vertices in $\G$.
The contributions of the two matrices to the model are as follows.
The combinatorial matrix $\mathbf\Xi$ encodes the configuration model as described above.
The propensity matrix $\mathbf\Omega$ encodes dyadic propensities of vertices that go beyond what prescribed by the combinatorial matrix $\mathbf\Xi$.
The ratio between any two elements $\Omega_{ij}$ and $\Omega_{kl}$ of the propensity matrix is the odds-ratio of observing an edge between vertices $i$ and $j$ instead $k$ and $l$, independently of the degrees of the vertices.

As for the case of the configuration model, this process can be seen as sampling edges from an urn.
Moreover, specifying a propensity matrix $\mathbf\Omega$ allows to bias the sampling in specified ways, so that some edges are more likely to be sampled than others.
The probability distribution over a graph $\G$ given $\mathbf\Xi$ and $\mathbf\Omega$ is then described by the multivariate Wallenius' non-central hypergeometric distribution~\cite{wallenius1963, Chesson1978}.

We further denote with $\mathbf A$ the adjacency matrix of the multi-graph $\G$ and with $V$ its set of vertices, the probability distribution underlying a gHypEG $\mathbb X(\mathbf\Xi,\mathbf\Omega,m)$ with parameters $\mathbf\Xi$, $\mathbf\Omega$, and with $m$ edges is defined as follows:
\begin{equation}
\label{eq:walleniusNet}
	\Pr(\G\lvert\mathbf\Xi,\mathbf\Omega)=\left[\prod_{i,j\in V}{\dbinom{\Xi_{ij}}{A_{ij}}}\right]
         \int_{0}^{1}{\prod_{i,j\in V}{\left(1-z^{\frac{\Omega_{ij}}{S_{\mathbf{\Omega}} }}\right)^{A_{ij}}}dz}
\end{equation}
with
\begin{equation}
	S_{\mathbf{\Omega}}= \sum_{i,j\in V} \Omega_{ij}(\Xi_{ij}-A_{ij}).
\end{equation}
The probability distribution for undirected graphs and for graphs without self-loops are defined similarly: by excluding the lower triangular entries of the adjacency matrix or by excluding its diagonal entries respectively (see \cite{Casiraghi2018} for more details).

In the case of large graphs, sampling from an urn without replacement can be approximated by a sampling with replacement from the same urn.
Under this assumption, the approximation allows to estimate the probability given in \cref{eq:walleniusNet} by means of a multinomial distribution with parameters $p_{ij}=\Xi_{ij}\Omega_{ij}/\sum_{kl}\Xi_{kl}\Omega_{kl}$.

\section{Block-constrained Configuration Model}\label{sec:blockghype}
\begin{figure}[h]
\centering
\includegraphics[width=.5\textwidth]{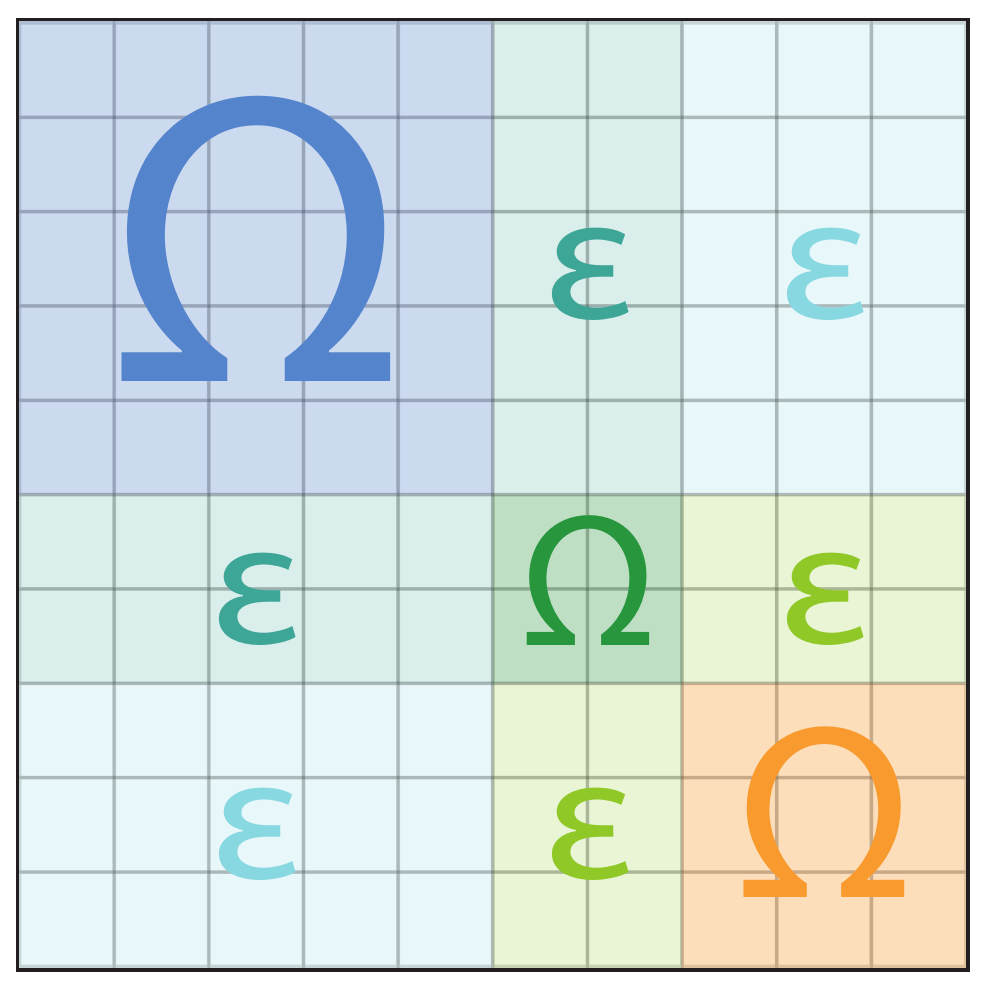}
\label{fig:blockomega}
\caption{\small Structure of a block propensity matrix with 3 different blocks (blue, green, yellow). The entries along the diagonal capture the within-block propensities, those away from the diagonal capture the between-block propensities.}
\end{figure}

Building on the framework provided by generalised hypergeometric ensembles of random graphs, we
define the block-constrained configuration model (BCCM) by means of a special form of the propensity matrix $\mathbf\Omega$.
Specifically, we need to encode the block structure that we observe in the propensity matrix $\mathbf\Omega$.
This is achieved by specifying a block propensity matrix $\mathbf\Omega^{(B)}$ where each of its elements ${\Omega^{(B)}}_{ij}=\omega_{b_i}$ if the vertices $i$ and $j$ are in the same block $b_i$, and ${\Omega^{(B)}}_{ij}=\omega_{b_ib_j}$ if the vertices $i$ and $j$ are in different blocks $b_i$ and $b_j$ respectively.
\Cref{fig:blockomega} shows a block-propensity matrix characterised by three blocks.
Similarly to the original SBM, in the presence of $B$ blocks we can specify a $B\times B$ block-matrix $\mathbf B$ that captures the block structure through its parameters $\omega_{b_ib_j}$.
However, in the case of a BCCM, the entries $\omega_{b_ib_j}$ capture the deviations in terms of edge propensities from the configuration model defined by the matrix $\mathbf\Xi$, constraining edges into blocks.

The block-matrix $\mathbf B$ can be specified to generate various structures, extending those naturally generated by degrees only, such as for instance a diagonal block-matrix can model graphs with disconnected components.
The inclusion of small off-diagonal elements gives rise to standard community structures, with densely connected clusters of vertices.
By specifying different types of block-matrices it is also possible to model core-periphery, hierarchical, or multipartite structures.

The block-constrained configuration model $\mathbb X(\mathbf\Xi,\mathbf B,m)$ with $m$ edges is thus completely defined by the combinatorial matrix $\mathbf\Xi$, and by the block-matrix $\mathbf B$ generating the propensity matrix $\mathbf\Omega^{(B)}$.
We can then rewrite the general probability for a gHypEG given in \cref{eq:walleniusNet} for BCCMs:
\begin{equation}
\label{eq:walleniusBCCM}
	\Pr(\G\lvert\mathbf\Xi,\mathbf B)=\left[\prod_{i,j\in V}{\dbinom{\Xi_{ij}}{A_{ij}}}\right]
         \int_{0}^{1}{\prod_{i,j\in V}{\left(1-z^{\frac{\omega_{b_ib_j}}{S_{\mathbf{B}} }}\right)^{A_{ij}}}dz}
\end{equation}
with
\begin{equation}
	S_{\mathbf{B}}= \sum_{i,j\in V} \omega_{b_ib_j}(\Xi_{ij}-A_{ij}).
\end{equation}
The analytical tractability provided by the closed-form solution of the distribution underlying BCCMs has two major advantages.

First, it allows to compute probabilities for large graphs, without the need to resort to Monte-Carlo simulations.
This permits the study of large graphs, and provides simple model selection methods based on the comparison of likelihoods, such as likelihood-ratio tests, or those based on information criteria.
In this article we will consider model selection based on the comparison of information criteria.
We will adopt the two most commonly used ones: Akaike information criterion (AIC)~\cite{Akaike1974}, and Schwarz or Bayesian information criterion (BIC)~\cite{schwarz1978estimating}.
Both criteria depend on the likelihood function of the models to be compared, and penalise for the number of parameters estimated by the model.
The model with the lowest score is the preferred one, as it best fits the data without overfitting it.

The Akaike information criterion for a model $\mathbb X$ given a graph $\G$ is formulated as follows:
\begin{equation}
	\AIC(\mathbb X|\G)=2k-2\log\left[\hat L(\mathbb X|\G)\right],
\end{equation}
where $k$ is the number of parameters estimated by $\mathbf X$ and $\hat L(\mathbb X|\G)$ is the likelihood of model $\mathbb X$ given the graph $\G$.

The Bayesian information criterion for a model $\mathbb X$ given a graph $\G$ is given by:
\begin{equation}
	\BIC(\mathbb X|\G)=\log(m)k-2\log\left[\hat L(\mathbb X|\G)\right],
\end{equation}
where $k$ is the number of parameters estimated by $\mathbb X$, $m$ is the number of observations, i.e., edges, and $\hat L(\mathbb X|\G)$ is the likelihood of model $\mathbb X$ given the graph $\G$.

The second major advantage given by the analytical tractability of the BCCM is the ability to easily estimate its block-matrix $\mathbf B$ from data.
Thanks to this we are able to fit BCCMs to large graphs without resorting to computationally expensive numerical simulations.

In the next sections, we describe how BCCMs can be used to generate graphs, and how to fit the block-matrix $\mathbf B$ to an observed graph.

\paragraph{Generating realisations from the BCCM.}
BCCMs are practical generative models that allow the creation of synthetic graphs with complex structures by drawing realisations from the multivariate Wallenius non-central hypergeometric distribution.
The process of generating synthetic graphs can be divided into two tasks.
First it is needed to specify the degree sequences for the vertices.
This can achieved by e.g. sampling the degree sequences from a power-law or exponential distributions.
From the degree sequences we can generate the combinatorial matrix $\mathbf\Xi$, specifying its elements $\Xi_{ij}=k_i^{\text{out}}*k_j^{\text{in}}$, where $k_i^{\text{out}}$ is the out-degree of vertex $i$.
Second, we need to define a block-matrix $\mathbf B$, whose elements define the propensities of observing edges between vertices, between and within the different blocks.

The block-matrix $\mathbf B$ takes the form given in \cref{eq:B}:
\begin{equation}\label{eq:B}
	\mathbf B= \begin{bmatrix}
    \omega_{b_1} & \dots & \omega_{b_1b_B} \\
    & \vdots & \\
    \omega_{b_Bb_1} & \dots & \omega_{b_B}
  \end{bmatrix}.
\end{equation}
Elements $\omega_{b_ib_j}$ should be specified such that the ratio between any two elements corresponds to the chosen odds-ratios of observing an edge in the block corresponding to the first element instead of the block corresponding to the second element, given the degrees of the corresponding vertices were the same.
For example, $\omega_{b_1}/\omega_{b_3b_2}$ corresponds to the odds-ratio of observing an edge between vertices in block 1 compared to an edge between block 2 and 3.
Note that in the case of an undirected graph, $\omega_{b_ib_j}=\omega_{b_jb_i}$ $\forall i,j$.
On the other hand, in the case of a directed graph blocks may have a preferred directionality, i.e., edges between blocks may be more likely in one direction.
In this case, we may choose $\omega_{b_ib_j}\neq\omega_{b_jb_i}$ for same pairs of vertices $i,j$.

Once the parameters of the model are defined, we sample graphs with $m$ edges from the BCCM $\mathbb X(\mathbf\Xi,\mathbf\Omega_B,m)$ defined by the combinatorial matrix $\mathbf\Xi$, and the block-propensity matrix $\mathbf\Omega_B$ defined by $\mathbf B$.
As described in the previous section, sampling a graph from $\mathbb X(\mathbf\Xi,\mathbf\Omega_B,m)$ corresponds to sample $m$ edges according to the multivariate Wallenius non-central hypergeometric distribution.
For example, this can be performed by means of the implementation \texttt{BiasedUrn} provided by~\citet{Fog2008a,Fog2008} in \texttt{C} and as a library for \texttt{R}.

\paragraph{Examples.}
We can specify different types of clustered graphs by means of this construction.
As demonstrative example\footnote{The code used to generate the examples described here, and that used for the case study analysis of the next section, can be found online at the url \url{https://github.com/gi0na/BCCM--Supporting-Material.git}.}, we define a block-matrix with 5 blocks connected in a ring.
Each block is as dense as the others, and blocks are weakly connected with only to their closest neighbours.
The block-matrix quantifying these specification is given as
\begin{equation}\label{eq:B5synt}
	\mathbf B= \begin{bmatrix}
    1 & 0.1 & 0 & 0 & 0 \\
    0 & 1 & 0.1 & 0 & 0 \\
    0 & 0 & 1 & 0.1 & 0 \\
    0 & 0 & 0 & 1 & 0.1 \\
    0.1 & 0 & 0 & 0 & 1 \\
  \end{bmatrix}.
\end{equation}
According to the choice made in \cref{eq:B5synt}, edges within diagonal blocks are 10 times more likely than edges within off-diagonal blocks.

\begin{figure}
	\centering
	\includegraphics[width=.33\textwidth]{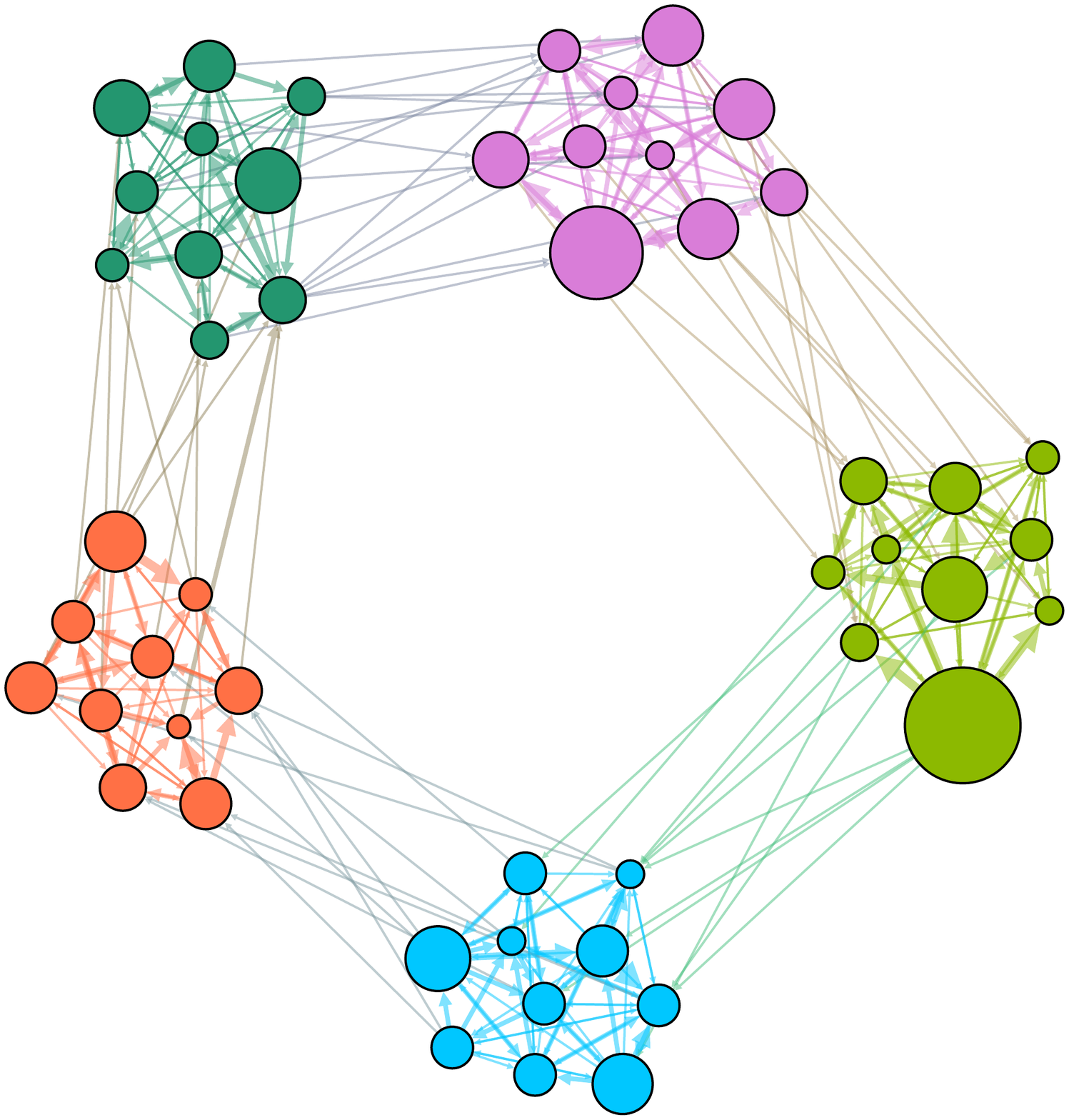}\hfill
	\includegraphics[width=.33\textwidth]{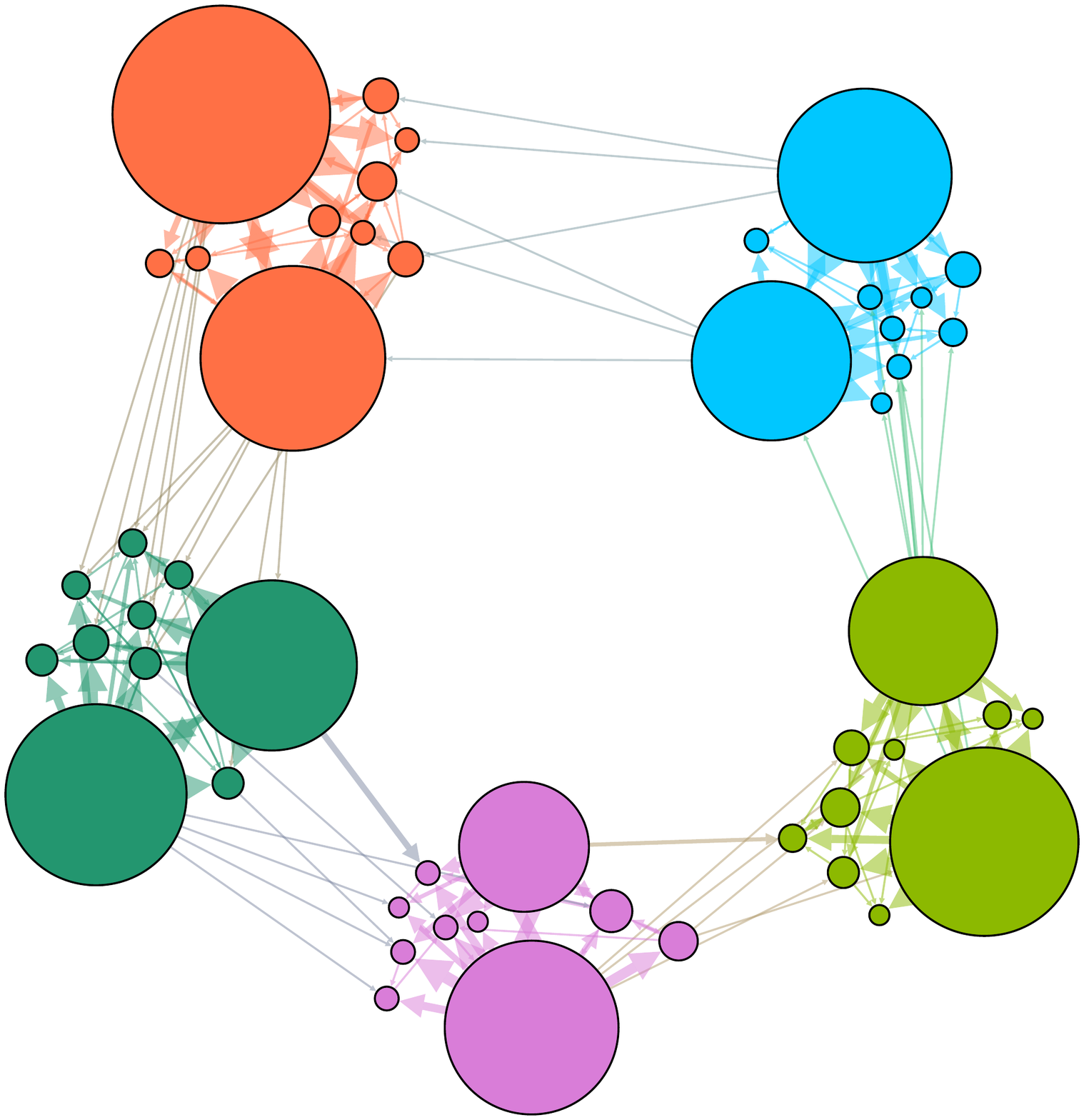}\hfill
	\includegraphics[width=.33\textwidth]{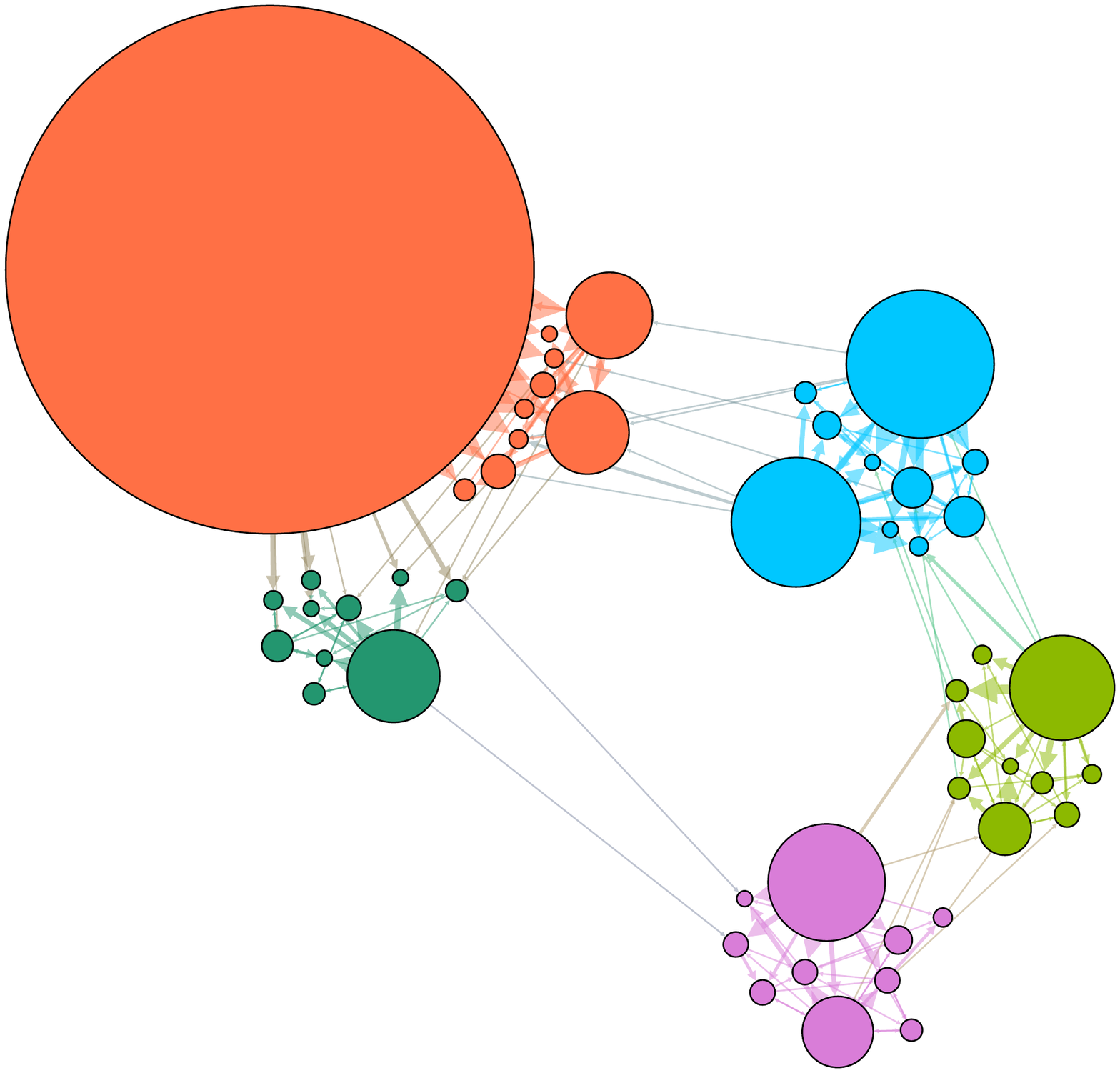}\\
	\hfill \textbf{(a)} \hfill \hspace{1em} \hfill \textbf{(b)} \hfill \vspace{1em} \hfill \textbf{(c)} \hfill \hspace{1em}
	\caption{\small 
	Realisations from a block-constrained configuration model obtained by fixing the block-matrix $\mathbf B$ and varying the out-degree distribution.
	Each realisation is obtained from a BCCM with $N=50$ vertices and $m=500$ directed edges.
	The vertices are separated into 5 equally sized blocks and the block-matrix $\mathbf B$ is given by \cref{eq:B5synt}.
	On left side, (a) is a realisation from a BCCM where the degree distributions are uniform.
	It corresponds to a realisation from a standard SBM.
	In the center, (b) is a realisation obtained by drawing the out-degree distribution of the vertices in each block from a power-law distribution with parameter $\alpha=1.8$.
	On the right side, (c) is a realisation obtained by drawing the out-degree distribution of all vertices from the same power-law.
	All graphs are visualised using the force-atlas2 layout with weighted edges.
	Out-degrees determine vertex sizes, and edge widths the edge counts.
	}\label{fig:examplering}
\end{figure}

After fixing this block-matrix, we can define different degree sequences for the vertices.
We highlight here the results obtained when fixing three different options in a directed graph without self-loops, with $n=50$ vertices and $m=500$ edges.
The first degree sequence we can set is the most simple option, corresponding to the standard non-degree-corrected stochastic block-model.
This model corresponds to setting each entry in the combinatorial matrix $\mathbf\Xi$ equal to $m^2/(n(n-1))$~\cite{Casiraghi2016}.
If assign the same number of vertices to each block, we expect the model to generate graphs with homogeneous blocks.
\Cref{fig:examplering} (a) shows a realisation from this model.
The second degree sequence we can set is defined such that the degrees of the vertices of each block are drawn from a power-law distribution.
We expect that each block shows the same structure, with few vertices with high degrees, and many with low degrees.
Because of this, we expect that most blocks are connected with directed edges starting from high-degree vertices.
\Cref{fig:examplering} (b) shows a realisation from this model where this clearly visible.
Finally, we set a degree sequence where the degrees of all vertices are drawn from a power-law distribution.
\Cref{fig:examplering} (c) shows a realisation from this model.

\begin{figure}
	\centering
	\includegraphics[width=.33\textwidth]{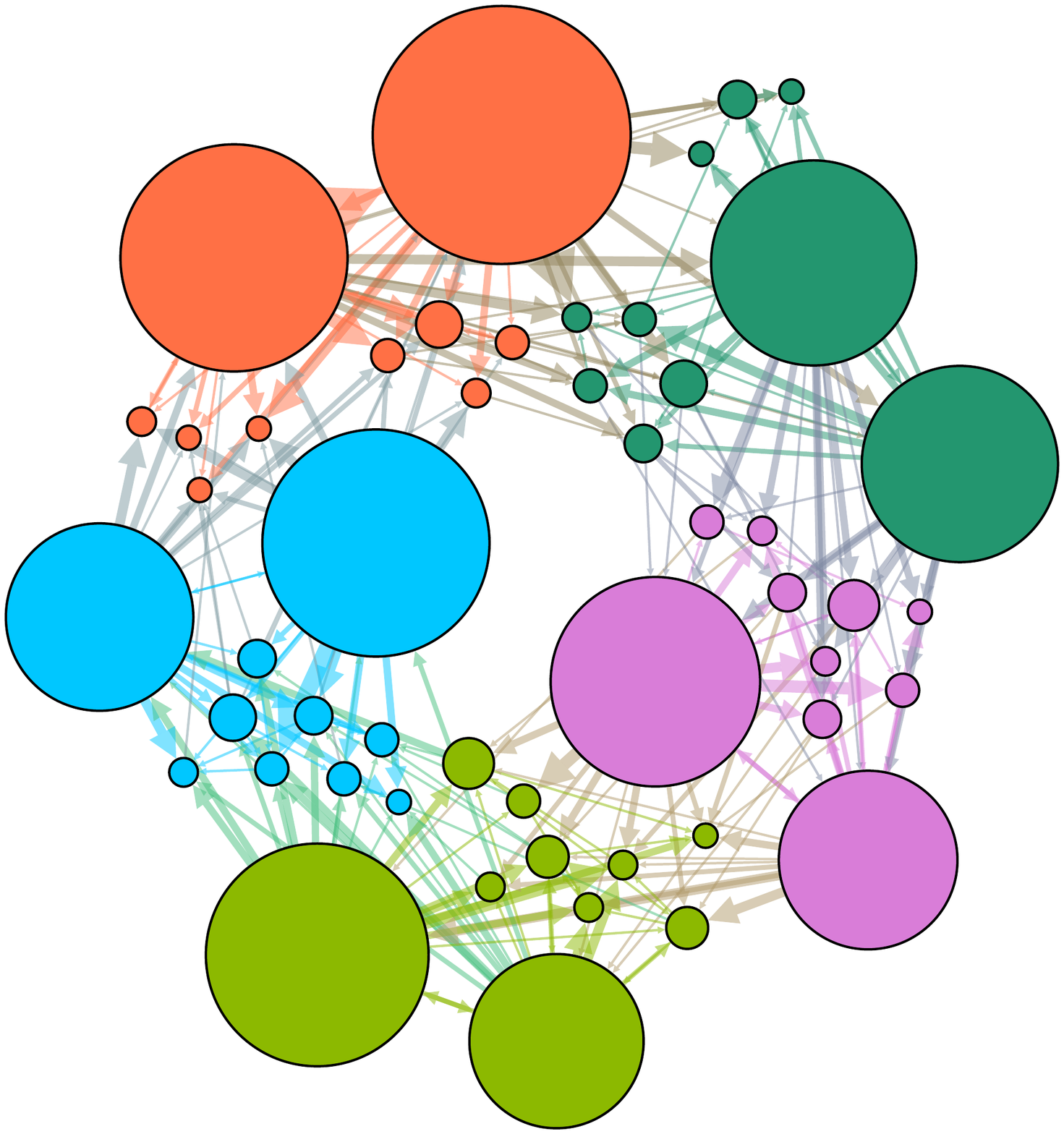}\hfill
	\includegraphics[width=.33\textwidth]{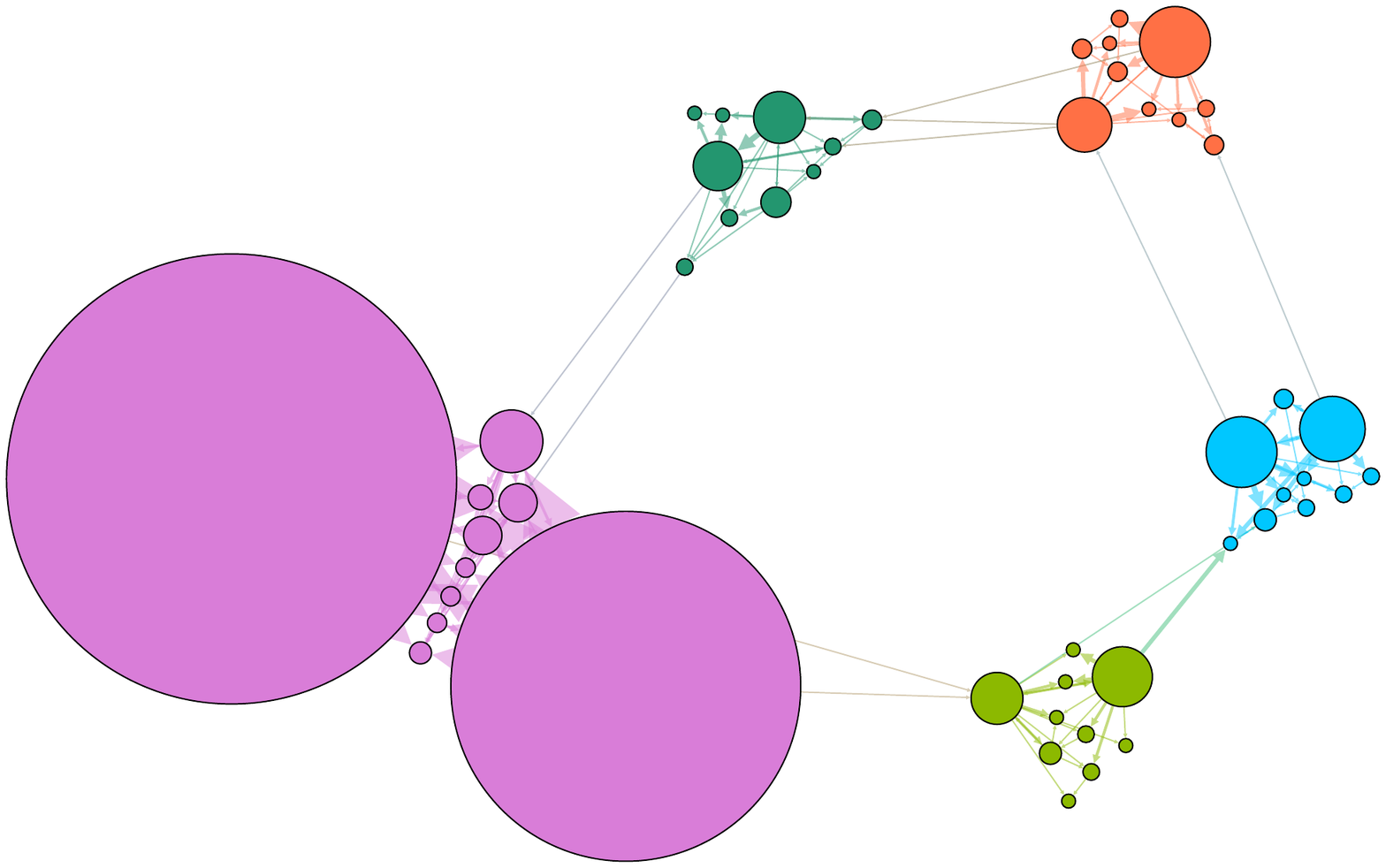}\hfill
	\includegraphics[width=.33\textwidth]{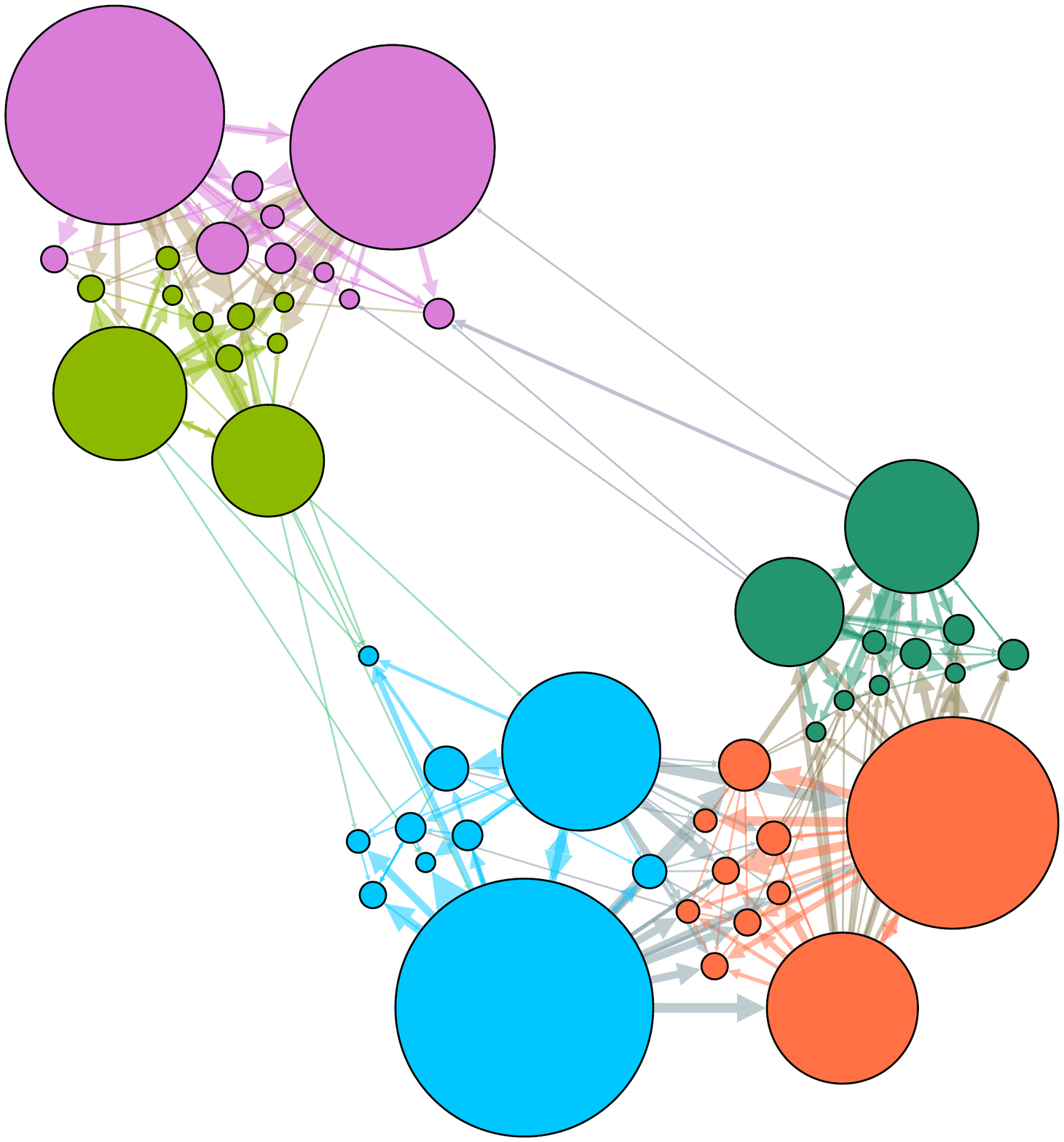}\\
	\hfill \textbf{(a)} \hfill \hspace{1em} \hfill \textbf{(b)} \hfill \vspace{1em} \hfill \textbf{(c)} \hfill \hspace{1em}
	\caption{\small 
		Realisations from a block-constrained configuration model obtained by fixing the out-degree distribution and varying the parameters within the block-matrix $\mathbf B$.
	Each realisation is obtained from a BCCM with $N=50$ vertices and $m=500$ directed edges.
	The out-degree distribution of the vertices in each block follows a power-law distribution with parameter $\alpha=1.8$
	The vertices are separated into 5 equally sized blocks and the structure of the block-matrix $\mathbf B$ is given by \cref{eq:B5synt}, but in each graph the values of some of the parameters $\omega_{b_ib_j}$ are changed.
	On left side, (a) is a realisation from a BCCM where the between-block parameters are increased to 1.
	In the center, (b) is a realisation obtained by increasing the parameter $\omega_{b_1}$ that controls for the internal cohesion of the first block.
	On the right side, (c) is a realisation obtained by increasing to 0.8 the between-block parameters $\omega_{b_1b_2}$, $\omega_{b_3b_4}$, and $\omega_{b_4b_5}$, to create a hierarchical block structure where the first two blocks are part of a macro cluster, and the last three blocks are part of another.
	All graphs are visualised using the force-atlas2 layout with weighted edges.
	Out-degrees determine vertex sizes, and edge widths the edge counts.
	}\label{fig:examplevarblock}
\end{figure}

Instead of varying the degree sequences of the underlying configuration model, we can as well vary the strength of block structure, changing the block-matrix $\mathbf B$.
Similarly to what we did above, we show three different combinations of parameters.
First, we set the within group parameters $\omega_{b_i}$ equal to the between group parameters $\omega_{b_ib_j}$ $\forall i,j$.
Second, we set the parameters $\omega_{b_1}=10$ so that the more edges are concentrated in the first block.
Third, we set the parameter to reconstruct a hierarchical structure.
We modify the parameters $\omega_{b_1b_2}=\omega_{b_3b_4}=\omega_{b_4b_5}=0.8$ to model graphs with two macro clusters weakly connected, where the one is split into two clusters strongly connected and the other into three clusters strongly connected.
Realisations drawn from each of these three models are shown in \cref{fig:examplevarblock}.

\paragraph{Fitting the block-matrix.}
The formulation of the block-constrained configuration model by means of the gHypEG framework allows for the fast estimation of the parameters of the block-matrix, in accordance with the graph that is being modelled.
Similarly to what is done with SBMs, we fit the BCCM by preserving in expectation the observed number of edges between and within different blocks.
To estimate the entries $\omega_b$ of the block-matrix $\mathbf{\hat{B}}$, we exploit the properties of the generalised hypergeometric ensemble of random graphs.

In gHypEs, the entries of the expected adjacency matrix \(\langle A_{ij}\rangle\) are obtained by solving the following system of equations~\cite{Casiraghi2018}:
\begin{align}\label{eq:walleniusMean}
\left(1-\frac{\langle A_{11}\rangle}{\Xi_{11}}\right)^{\frac{1}{\Omega_{11}}} = \left(1-\frac{\langle A_{12}\rangle}{\Xi_{12}}\right)^{\frac{1}{\Omega_{12}}} = \ldots \end{align}
with the constraint $\sum_{i,j \in V} \langle A_{ij}\rangle = m$.

Because to estimate BCCMs we need to fix the expectation of the number of edges between blocks and not between dyads, we proceed as described below.
We denote with $A_{b_\alpha}=\sum_{i,j\in b_\alpha}A_{ij}$ the number of edges between all vertices $i,j$ that are in the same block $b_\alpha$, and similarly with $\Xi_{b_\alpha}=\sum_{i,j\in b_\alpha}\Xi_{ij}$ the sum of all the elements of the matrix $\mathbf\Xi$ corresponding to those dyads.
Then, we fix the expectations of the ensemble such that the number of edges between and within blocks are given by $A_{b_\alpha}$s.
Hence, in the case of the block-constrained configuration model with $B$ blocks we estimate the $B\cdot (B+1)/2$ parameters $\omega_{b_\alpha b_\beta}$s constituting the block-matrix $\mathbf{\hat{B}}$ solving the following set of independent equations, defined up to an arbitrary constant $k$:
\begin{align}\label{eq:walleniusBlock}
	\begin{cases}
		\left(1-\frac{A_{b_1}}{\Xi_{b_1}}\right)^{\frac{1}{\omega_{b_1}}} &= k\\
		&\vdots\\
		\left(1-\frac{A_{b_B}}{\Xi_{b_B}}\right)^{\frac{1}{\omega_{b_B}}} &= k\,.
	\end{cases}
\end{align}
Solving for $\omega_{b_\alpha b_\beta}$, we find that the entries of the block-matrix $\mathbf{\hat{B}}$ that preserve in expectation the observed number of edges between and within blocks are given by
\begin{equation}
	\omega_{b_\alpha b_\beta} := -\log(1-\frac{A_{b_\alpha b_\beta}}{\Xi_{b_\alpha b_\beta}}).
\end{equation}
The estimation of the parameters scales quadratically only with the number of blocks.
It is hence simple to fit the parameters of BCCMs with fixed block structure even for large graphs.

When the parameters of the BCCM are estimated as described here, the block-constrained configuration model has the advantageous property of asymptotic consistency. This means that if the method described here is applied to synthetic graphs generated from a BCCM, the method introduced in this article can correctly recover the original model.

\section{Case Studies}\label{sec:casestudy}

We conclude the article with a case study analysis of synthetic and empirical graphs.
We highlight the interpretability of the resulting block-constrained configuration models in terms of deviations from the classical configuration model.
In particular, a weak community structure in a graph is reflected in a small contribution to the likelihood of the estimated block-matrix.
On the other hand, a strong community structure is reflected by a large contribution to the likelihood by the estimated block-matrix.
Here, we quantify this difference by means of AIC or BIC.
However, other information criteria may also be used.
Moreover, studying the relative values of the estimated parameters in the block matrices quantifies how much the configuration model has to be biased towards a block structure to optimally fit the observed graph.
The more different are the values of the parameters, the stronger is the block structure compared to what is expected from the configuration model.

We start by analysing synthetic graphs generated according to different rules, and we show that fitting the block-constrained configuration model parameters allows to select the correct, i.e., planted, partition of vertices, among a given set of different partitions.
We perform three experiments with large directed graphs with clusters of different sizes.
Finally, we conclude by employing the BCCM to compare how well different partitions obtained by means of different clustering algorithm fit well-known real world networks.

\paragraph{Analysis of synthetic graphs.}
We generate synthetic graphs incorporating `activities' of vertices in a classical SBM, to be able to plant different out-degree sequences in the synthetic graphs. 
First, we need to assign the given activity to each vertex. Higher activity means that the vertex is more likely to have a higher degree.
Second, we need to assign vertices to blocks, and assign a probability of sampling edges to each block.
Densely connected blocks have a higher probability than weakly connected blocks.
The graph is then generated by a weighted sampling of edges with replacement from the list containing all dyads of the graph.
Weights to sample each dyad are given by the product between the activity corresponding to the from-vertex, and the weight corresponding to the block to which the dyad belongs.
The probabilities of sampling edges correspond to the normalised weights, so that their sum is 1.

For example, let's assume we want to generate a 3 vertices graphs with two clusters.
We can fix the block weights as follows: edges in block 1 or 2 have weight $w_1$ and $w_2$ respectively; edges between block 1 and block 2 have weight $w_{12}$.
\Cref{eq:tabcreation} shows the list of dyads from which to sample together with their weights, where the activity of vertices is fixed to $(a_1,a_2,a_3)$, and the first two vertices belong to the first block.
\begin{table}
\centering
	\begin{tabular}{c|ccc|c}
		\text{dyad} & \text{activity} & \text{block id} & \text{block weight} & \text{sampling weight}\\
		\hline
		1 -- 1 & $a_1$  & 1  & $w_1$    & $a_1w_1$ \\
		1 -- 2 & $a_1$  & 1  & $w_1$    & $a_1w_1$ \\
		1 -- 3 & $a_1$  & 12 & $w_{12}$ & $a_1w_{12}$ \\
		2 -- 1 & $a_2$  & 1  & $w_1$    & $a_2w_1$ \\
		2 -- 2 & $a_2$  & 1  & $w_1$    & $a_2w_1$ \\
		2 -- 3 & $a_2$  & 12 & $w_{12}$ & $a_2w_{12}$ \\
		3 -- 1 & $a_3$  & 12 & $w_{12}$ & $a_3w_{12}$ \\
		3 -- 2 & $a_3$  & 12 & $w_{12}$ & $a_3w_{12}$ \\
		3 -- 3 & $a_3$  & 2  & $w_2$    & $a_3w_2$ \\
	\end{tabular}
	\caption{
	Edge list with weights for the generation of synthetic graphs with given vertex activities and block structure.
	}\label{eq:tabcreation}
\end{table}
Note that if the activities of the vertices were all set to the same value, this process would correspond to the original SBM.
In the following experiments, we generate different directed graphs with $N=500$ vertices, $m=40000$ edges, and different planted block structures and vertex activities.

In the first experiment, we show the difference between estimating the parameters for an SBM and for the BCCM when the block structure is given.
To do so, we first generate the activities of vertices from an exponential distribution with parameter $\lambda=N/m$ (such that the expected sum of all activities is equal to the number of edges $m$ we want to sample).
After sorting the activity vector in decreasing order, we assign it to the vertices.
In this way the first vertex has the highest activity, and hence highest out-degree, and so on.
In this first experiment we do not assign block weights so that the graphs obtained do not show any consistent cluster structure, and have a skewed out-degree distribution according to the fixed vertex activity (correlation $\sim1$).

First, we assign the vertices randomly to two blocks.
We proceed by estimating the parameters for an SBM and a BCCM, according to the blocks to which the vertex have been assigned.
Since no block structure has been enforced and the vertex have been assigned randomly to blocks, we expect that the estimated parameters for the block matrices $\mathbf{\hat B}_{\text{SBM}}$ and $\mathbf{\hat B}_{\text{BCCM}}$ will all be close to 1\footnote{When normalised by the maximum value.}, reflecting the absence of a block structure.
The resulting estimated parameters for an exemplary realisation are reported in \cref{eq:Bsynt1}.
\begin{equation}\label{eq:Bsynt1}
	\mathbf{\hat B}_{\text{SBM}}= \begin{bmatrix}
    1.0000000 & 0.9992577\\
	0.9992577 & 0.9603127
  \end{bmatrix}
  \hfill\quad
  	\mathbf{\hat B}_{\text{BCCM}}= \begin{bmatrix}
    0.9808935 & 1.0000000\\
	1.0000000 & 0.9805065
  \end{bmatrix}
\end{equation}
As expected, the estimated values for both models are close to 1.

After changing the way vertices are assigned to blocks, we repeat the estimation of the two models.
Now, we separate the vertices into two blocks such that the first $250$ vertices ordered by activity are assigned to the first block and the last $250$ to the second one.
We expect that the SBM will assign different parameters to the different blocks, because now the first block contains all vertices with high degree, and the second block all vertices with low degree.
Hence, most of the edges are found between vertices in the first block or between the two blocks.
Differently from the SBM, the BCCM corrects for the observed degrees.
Hence, we expect that the parameters found for the block-matrix will be all close to 1 again, as no structure beyond that one generated by the degrees is present.
Thus the block assignment does not matter for the estimated parameter.
The block matrices for the two models, estimated for the same realisation used above, are provided in \cref{eq:estB2}.
\begin{equation}\label{eq:estB2}
	\mathbf{\hat B}_{\text{SBM}}= \begin{bmatrix}
    1.000000 & 0.597866\\
	0.597866 & 0.194896
  \end{bmatrix}
  \hfill\quad
  	\mathbf{\hat B}_{\text{BCCM}}= \begin{bmatrix}
    0.997024 & 0.995108\\
	0.995108 & 1.000000
  \end{bmatrix}
\end{equation}
We observe that the SBM assigns different values to each block, impairing the interpretability of the result.
In particular, the parameters of $\mathbf{\hat B}_{\text{SBM}}$ show the presence of a core-periphery structure which cannot be distinguished from what obtained naturally from a skewed degree distributions.
The estimation of $\mathbf{\hat B}_{\text{BCCM}}$, on the contrary, highlights the absence of any block structure beyond that one generated by the degree sequence, and we can correctly conclude that the core-periphery structure of the observed graph is entirely generated by the degree distributions.

In the second synthetic experiment we highlight the model selection features of the BCCM.
Thanks to the fact that we are able to compute directly the likelihood of the model, we can easily compute information criteria such as AIC or BIC to perform model selection.
We generate directed graphs with self-loops with $N=500$ vertices, $m=40000$ edges, and 2 equally sized clusters.
Again, we generate vertex activities from an exponential distribution with rate $\lambda=N/m$.
We fix the block weights to be $w_1 = 1$, $w_2 = 3$, and $w_{12} = 0.1$.
By means of this setup we are able to generate synthetic graphs with two clusters, one of which is denser than the other.
If we fit a BCCM to the synthetic graph with the correct assignment of vertices to blocks we obtain the following block-matrix $\mathbf{\hat B}_{\text{BCCM}}$ for an exemplary realisation:
\begin{equation}
	\mathbf{\hat B}_{\text{BCCM}}= \begin{bmatrix}
    1.1760878 & 0.1108463\\
	0.1108463 & 3.0000000
  \end{bmatrix}
\end{equation}
We note that we approximately recover the original block weights used to generate the graph.

We can now compare the AIC obtained for the fitted BCCM model, $\AIC_{\text{BCCM}} = 662060$, to that obtained from a simple configuration model (CM) with no block assignment, $\AIC_{\text{CM}} = 693540$.
The CM model is formulated in terms of a gHypEG where the propensity matrix $\mathbf\Omega\equiv1$.
The AIC for the BCCM is considerably smaller, confirming that the model with block structure fits better the observed graph.
As benchmark, we compute the AIC for BCCM models where the vertices have been assigned randomly to the two blocks.
\Cref{eq:aicboot} reports the AICs obtained for 1000 random assignment of vertices to the blocks, computed on the same observed graph.
\begin{equation}\label{eq:aicboot}
\AIC = \begin{tabular}{cccccc}
	Min. & 1st Qu. & Median  &  Mean & 3rd Qu. & Max. \\
 693531  & 693543 &  693544 & 693543 & 693544 & 693544 
\end{tabular}	
\end{equation}
We observe that this usually results in values close to that of the simple configuration model, as the block assignment do not reflect the structure of the graph.
In few cases, a small number of vertices is correctly assigned to blocks, showing a small reduction in AIC, which is however far from that of the correct assignment.

BCCMs allow also to compare models with different number of blocks.
To do so we separate the vertices in one of the blocks of the model above into two new blocks.
Because we add more degrees of freedom, we expect an increase in the likelihood of the new BCCM with three blocks, but this should not be enough to give a considerable decrease in AIC.
In fact, since the synthetic graph has been built planting two blocks, the AIC should allow us to select as optimal model the BCCM with two blocks.
The resulting block-matrix $\mathbf{\hat B}_{\text{BCCM}}^{(3)}$ with three blocks is reported in \cref{eq:B3}.
\begin{equation}\label{eq:B3}
	\mathbf{\hat B}_{\text{BCCM}}^{(3)}= \begin{bmatrix}
    1.1739475 & 1.1797875 & 0.1088987\\
	1.1797875 & 1.1706410 & 0.1129094\\
	0.1088987 & 0.1129094 & 3.0000000
  \end{bmatrix}
\end{equation}
We see that the estimated model fits different parameter values for the two sub-blocks, since the added parameters can now accommodate for random variations generated by the edge sampling process.
However, as expected, there is no (statistical) evidence to support the more complex model. 
In fact, comparing the AIC values we obtain $\AIC_{\text{BCCM}}^{(3)} = 662065 > 662060 = \AIC_{\text{BCCM}}$.
This shows that we can successfully use BCCM to perform model selection, both when different number of clusters or different vertex assignments are used. 

\begin{figure}
\centering
\includegraphics[width=.5\textwidth]{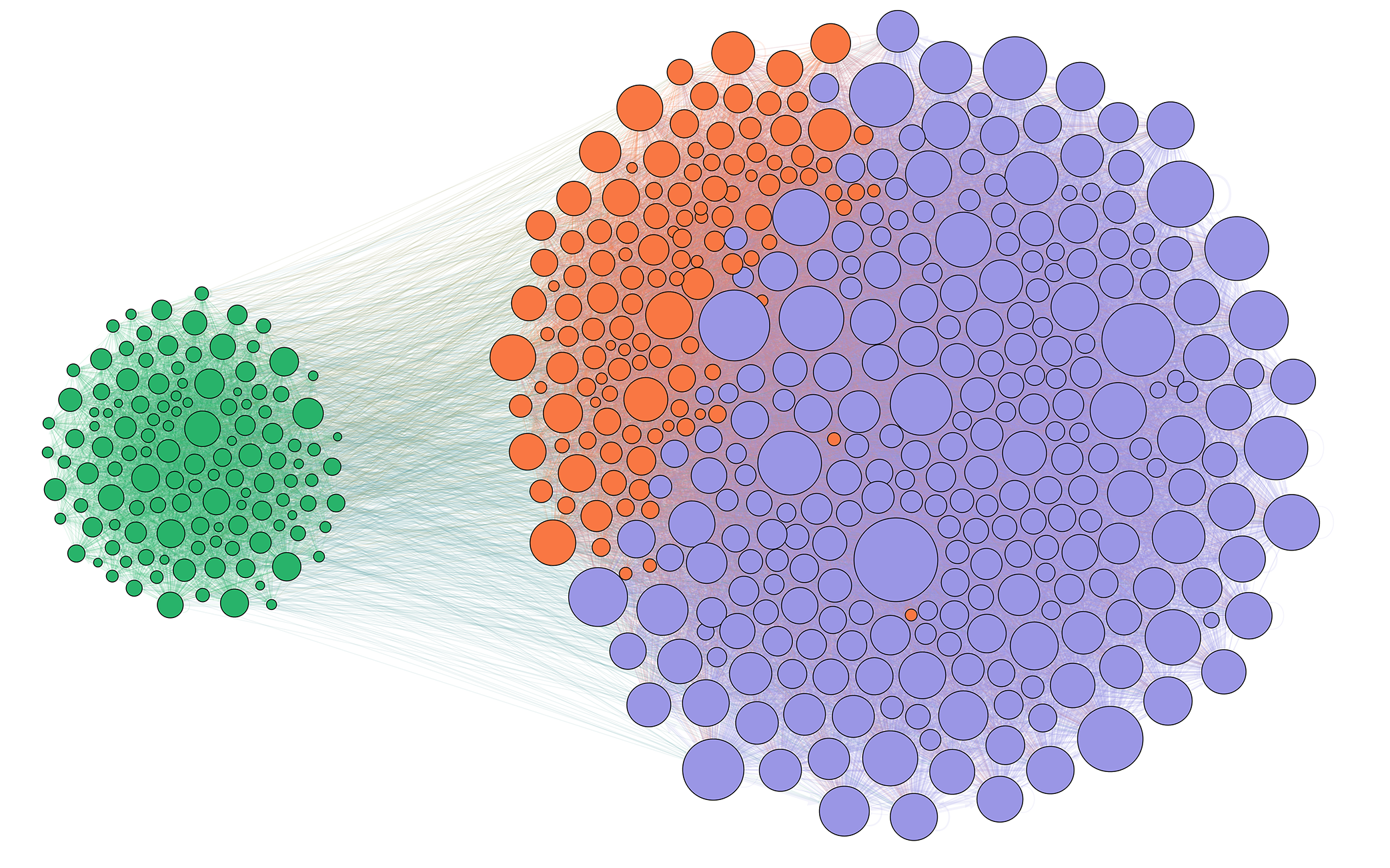}
\caption{\small 
Visualisation of a synthetic graph with $N=500$ vertices and $m=40000$~directed edges, obtained with the force-atlas2 layout.
Vertices are separated into three blocks of different sizes, such that the largest block (250 vertices, in purple) is strongly connected with one of the smaller blocks (125 vertices, in orange).
Both blocks are weakly connected to the third block, that is clearly separated (125 vertices, in green).
The out-degree sequence of the graph follows an exponential distribution with parameter $\lambda=N/m$.
The joint effects of the non-uniform degree sequence together with the asymmetric block structure makes the task of community detection on this graph particularly hard for standard algorithms.
}\label{fig:synt3}	
\end{figure}

In the third experiment, instead of two clusters, we plant three clusters of different sizes $(\abs{B_1}=250,\, \abs{B_2}=125,\,\abs{B_3} = 125)$.
We choose the block parameters such that one of the smaller cluster is more densely connected with the bigger cluster, and the smaller cluster is relatively more dense than the others.
To do so we choose the block weights as follows: $w_1 = w_2 = 1$, $w_3 = 3$, $w_{13} = w_{23} = 0.1$, $w_{12} = 0.8$.
As before, we draw vertex activities from an exponential distribution with parameter $\lambda=N/m$.
One exemplary realisation is plotted in \cref{fig:synt3}.
The plot clearly shows the separation into three clusters, with cluster 1 (purple) and 2 (orange) more densely connected to each other than to cluster 3 (green).
Fitting the same BCCMs as before allows to compare the AICs for the three-blocks BCCM to the 2-block BCCM.
In this case we expect that the model with 3 blocks will fit considerably better the graph.
Results of the fitting for the realisation plotted in \cref{fig:synt3} give $\AIC_{\text{BCCM}}^{(3)} = 673585 < 699765 = \AIC_{\text{BCCM}}^{(2)}$, correctly selecting the more complex model.
It is known that AIC does not punish model complexity as much as BIC.
For this reason, in this case we compare also the values of BIC obtained for the two models.
Also in this case, with $\BIC_{\text{BCCM}}^{(3)} = 2822787 < 2848941 = \BIC_{\text{BCCM}}^{(2)}$, the information criterion allows to correctly select the model with 3 blocks.

Finally, we can use AIC and BIC to evaluate and rank the goodness-of-fit different block assignments that are obtained from various community detection algorithms.
This allows to choose the best block assignment in terms of deviations from the configuration model, i.e., which of the detected block assignment better captures the block structure that go beyond that generated by the degree sequence of the observed graph.
We compare the result obtained from 5 different algorithms run using their \texttt{igraph} implementation for \texttt{R}.
In the following we use: \texttt{cluster\_fast\_greedy}, a greedy optimisation of modularity~\cite{clauset2004finding}; \texttt{cluster\_infomap}, the implementation of \texttt{infomap} available through \texttt{igraph}~\cite{rosvall2008maps}; \texttt{cluster\_label\_prop}, label propagation algorithm~\cite{raghavan2007near}; \texttt{cluster\_spinglass},  find communities in graphs via a spin-glass model and simulated annealing~\cite{reichardt2006statistical}; \emph{cluster\_louvain}, the Louvain multi-level modularity optimisation algorithm~\cite{blondel2008fast}.
As the modularity maximisation algorithms are implemented only for undirected graphs, we apply them to the undirected version of the observed graph.
The results of the application of the 5 different algorithms on the realisation shown in \cref{fig:synt3} are reported in the table in \cref{eq:aicsynt}.
\begin{table}
\centering
	\begin{tabular}{r|ccccc|c}
	 & fast\_greedy  & infomap  &  label\_prop & spinglass  & louvain & original \\
	\hline
	B &  2     &    4     &    2      &   7    &     2 & 3 \\
	AIC & 673871 &  \textbf{673867} &  673871 &  673907 &  673871 & 673585 \\
	BIC & \textbf{2823047} &  2823104 &  \textbf{2823047} &  2823298  & \textbf{2823047} & 2822787
\end{tabular}
\caption{Comparison of the goodness-of-fit of 5 different block structures detected by 5 different community detection algorithms.
The different partitions are compared in terms of the AIC and BIC obtained by the corresponding BCCM.
On the right-most column, are given the results corresponding to the ground-truth block partitioning.
}\label{eq:aicsynt}
\end{table}
The five different community detection algorithms find three different block structures.
Three of them are not able to detect the third block, while the other two algorithms split the vertices into two many blocks.
AIC ranks best \texttt{infomap} even though it detects one block too many.
BIC punishes for the number of parameters more, so ranks best the 2-blocks.
These results are consistent when repeating the experiment with different synthetic graphs generated from the same model.
It is worth noting that none of the community detection algorithms was able to correctly detect the planted block structure.
However, both the AIC and BIC of the BCCM fitted with the correct block structure are lower than those found by the different algorithms.
This shows that information criteria computed using BCCMs have a potential to develop novel community detection algorithms that are particularly suited for applications where degree correction is crucial.
However, the development of such algorithms is beyond the scope of this article and is left to future investigations.

\paragraph{Analysis of empirical graphs}
\begin{figure}
\centering
\includegraphics[width=.25\textwidth]{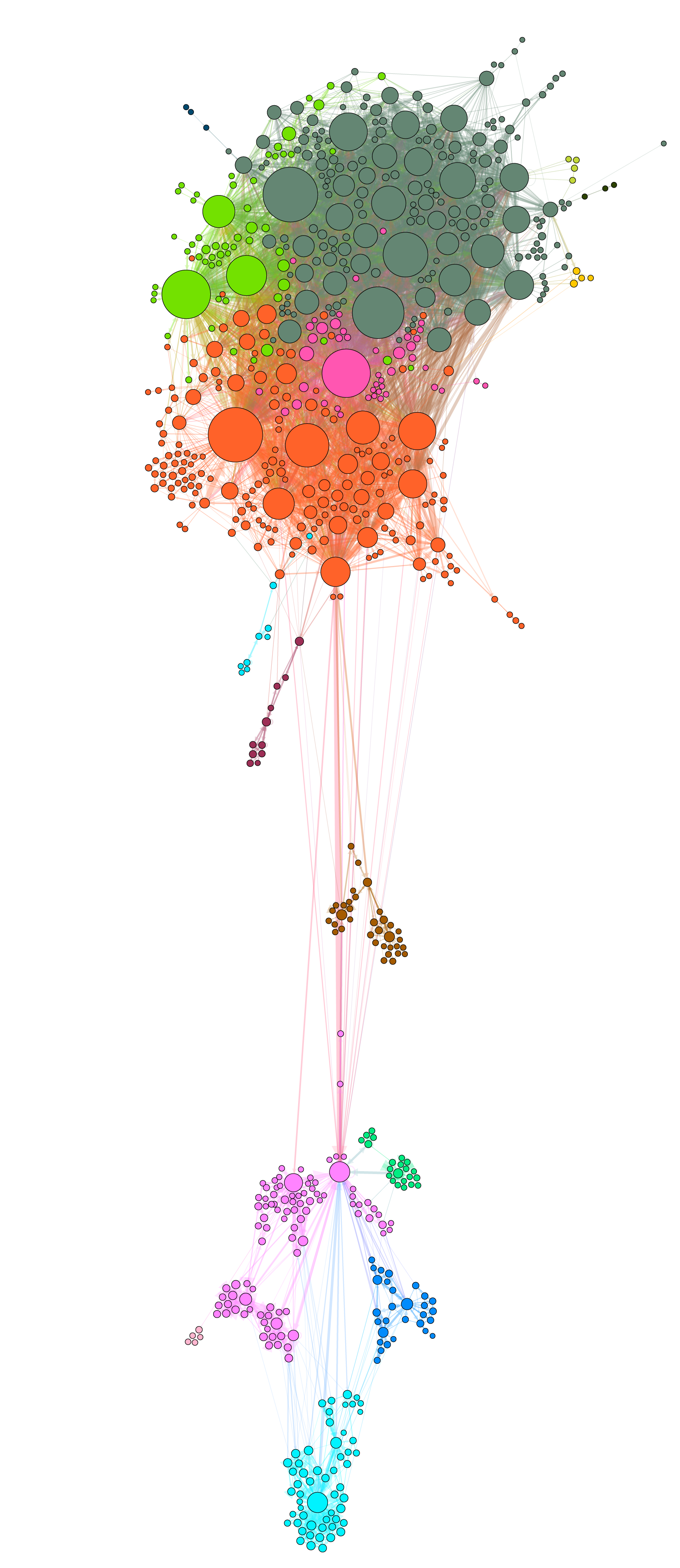}\hfill
\includegraphics[width=.25\textwidth]{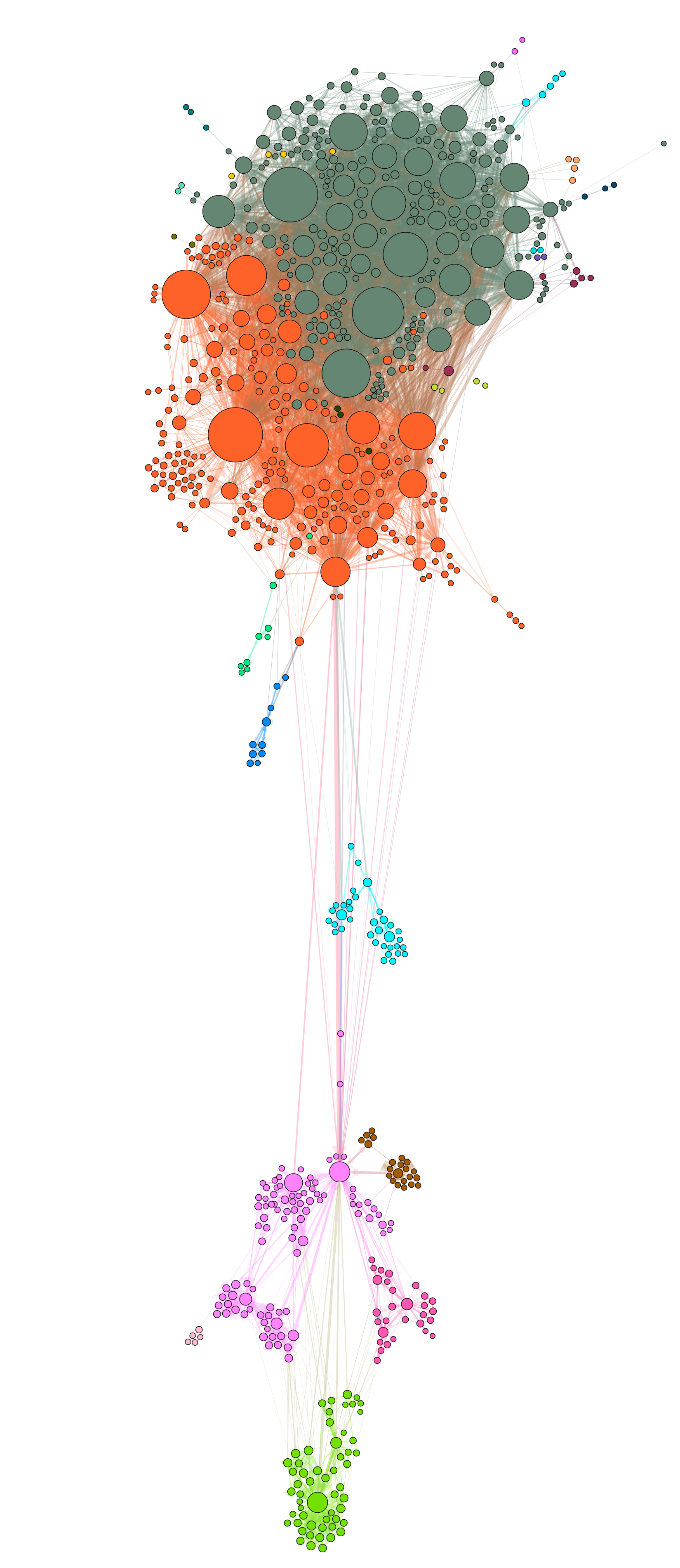}\hfill
\includegraphics[width=.25\textwidth]{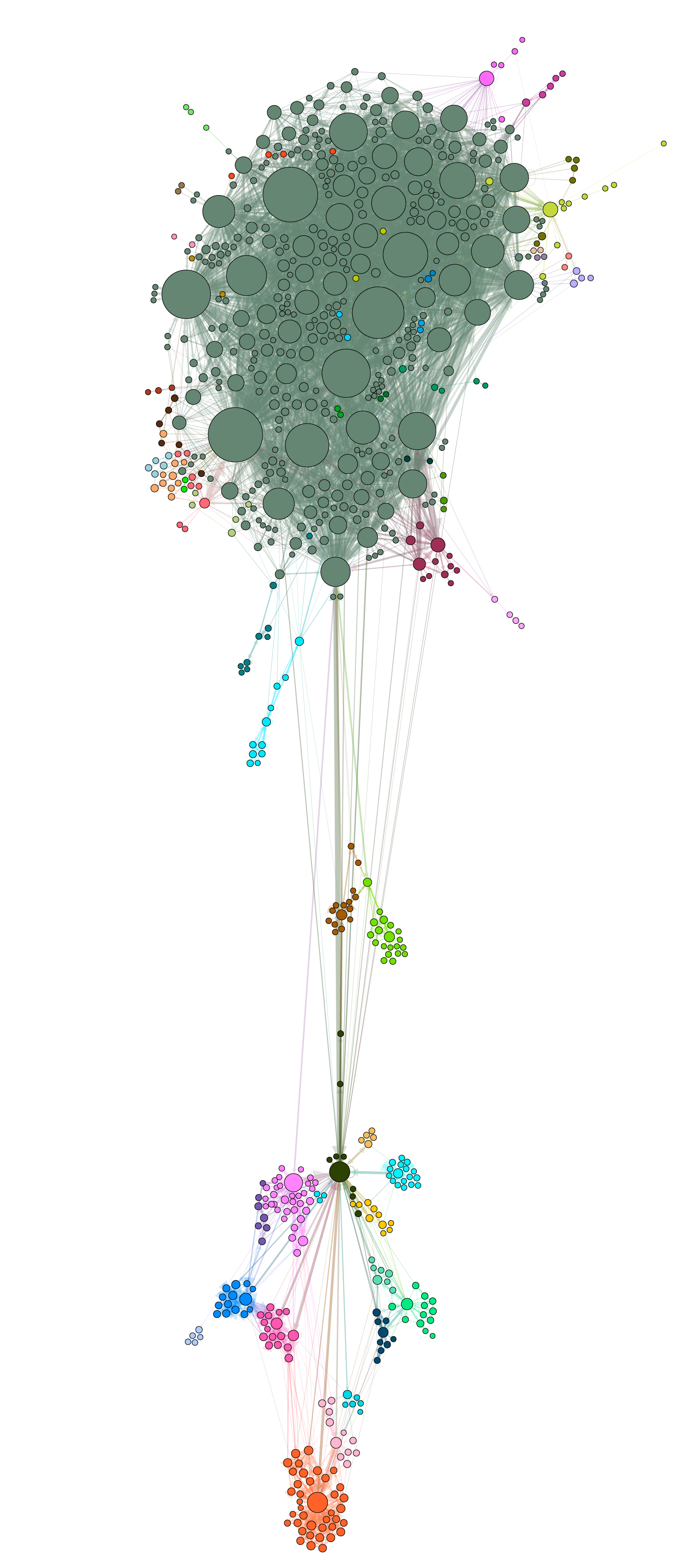}\hfill
\includegraphics[width=.25\textwidth]{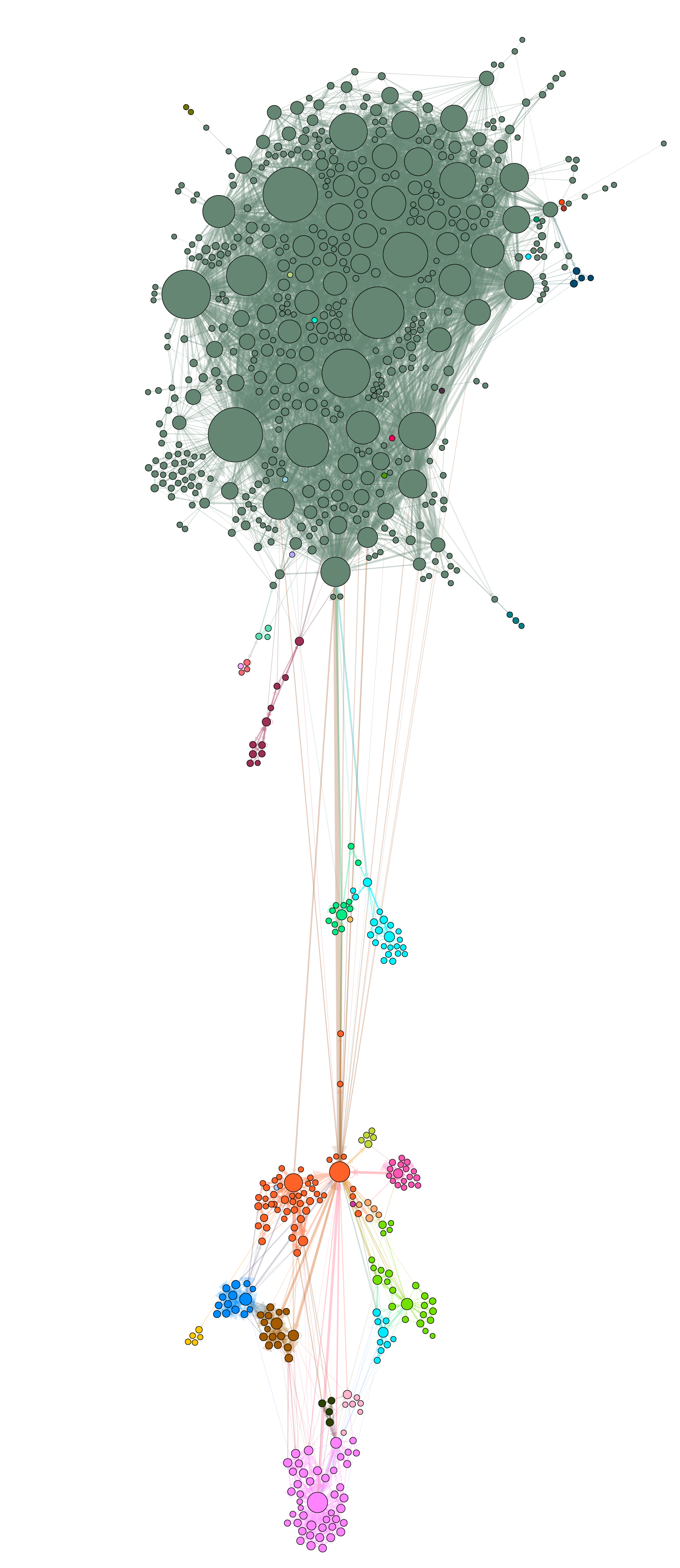}\\
\hspace{1em} \hfill \textbf{(louvain)} \hfill \hspace{1em} \hfill \textbf{(fast\_greedy)} \hfill \vspace{1em} \hfill \textbf{(infomap)} \hfill \hspace{1em} \hfill \textbf{(label\_prop)} \hfill \hspace{1em}
\caption{\small 
	\texttt{USairports} graph visualisation. The graph is plotted by means of the force-atlas2 layout with weighted edges, and the size of the vertices reflects their out-degrees.
	Only the largest connected component of the graph is shown.
	The visualisations clearly show the block structure that characterises this graph.
	The vertices in the four visualisations are coloured according to the labels detected applying four community detection algorithms, as described in \cref{tab:emp}.
	The visualisations are ordered from left to right according to the AIC of the BCCM fitted to observed graph according to the corresponding block structure.
	From left to right, we see the colours corresponding to the labels obtained from louvain, fast\_greedy, infomap and label\_propagation detection algorithms respectively.
	We highlight the fact that the ranking according to AIC corresponds approximately to the ability of the algorithms to detect the separation between high-degree (and low-degree) vertices within the largest cluster, at the top of the visualisations.
	The reason for this is that within the largest cluster there are clear deviations from what the configuration model predicts, i.e., high-degree vertices tend to connect to each other, and the best BCCMs captures more of these deviations.
	}\label{fig:airport}
\end{figure}
We conclude this article providing a comparison of the BCCMs obtained by fitting the block structures detected by the five community detection algorithms described above on five different real world networks.
The results show that different algorithm performs better for different graphs, highlighting the non-trivial effect that degrees have on block structure and community detection in general.

We study five well-known graphs with heterogeneous characteristics and sizes.
All graphs are multi-edge, and are freely available as dataset within the \texttt{igraphdata} \texttt{R} package. 
The first graph analysed is \texttt{rfid}: hospital encounter network data.
It consists of 32424 undirected edges between 75 individuals~\cite{vanhems2013estimating}.
The second graph analysed is \texttt{karate}: Zachary's Karate Club.
It consists of 231 undirected edges between 34 vertices~\cite{Zachary}.
The third graph analysed is \texttt{UKfaculty}: Friendship network of a UK university faculty.
It consists of 3730 directed edges between 81 vertices~\cite{nepusz2008fuzzy}.
The fourth graph is \texttt{USairports}: US airport network of December 2010.
It consists of 23473 directed edges between 755 airports~\cite{von2002comparative}. It has self-loops.
The graph is plotted in \cref{fig:airport}, using the force-atlas2 layout~\cite{jacomy2014forceatlas2}.
The four different plots are coloured according to the block structures detected by four of the five algorithms (\texttt{cluster\_spinglass} cannot be applied as the graph is disconnected).
They are ordered by increasing AIC.
From the visualisation we can see that best block structure is the one which is able to separate three different blocks within the largest cluster of vertices (top of the visualisations).
In particular, it is important to note that the largest cluster consist of high- and low-degree vertices.
If these vertices belonged to the same block, the configuration model predicts then high-degree vertices should be connected by many edges (similarly to the first synthetic experiment described above).
However, we observe then some of these high-degree vertices are separated and mainly connected to low-degree vertices.
For this reason, block structures that are able to separate these high-degree vertices into different blocks rank higher than others.
The fifth graph analysed is \texttt{enron}: Enron Email Network.
It consists of 125409 directed edges between 184 individuals~\cite{priebe2005scan}. It has self-loops.

Each of these graphs has a clear block structure that could be detected.
The different algorithms provide different results, both in the number of blocks detected and in the assignment of vertices.
Ranking the different results by means of the goodness-of-fit of BCCMs fitted according to the different block partitions shows that the best results are not necessarily those with fewer or more blocks, nor those obtained from a specific algorithm, as the results change with the graph studied.
The results of this analysis are provided in \cref{tab:emp}, where the smallest AICs and BICs for each graph are highlighted in bold, together with the algorithm that provides the smallest number of blocks.
The algorithm that provides the largest number of blocks is highlighted in italic.

\begin{table}
\small
\centering
\begin{tabular}{r|cccc}
	\multicolumn{5}{ c }{\textbf{Data Specifications}} \\
	dataset & vertices  & edges  &  directed & self-loops \\
	\hline
	rfid &  75     &    32424    &     False    &     False   \\
	karate & 34     &    231   &      False   &     False \\
	UKfaculty & 81    &    3730    &     True     &    False \\
	USairports & 755     &   23473      &  True   &      True\\
	enron & 184    &   125409    &    True    &     True  \\

\end{tabular}
\bigskip

\begin{tabular}{r|ccccc}
	\multicolumn{6}{ c }{\textbf{Number of Clusters}} \\
	dataset & fast\_greedy  & infomap  &  label\_prop & spinglass  & louvain \\
	\hline
	rfid &  6     &    4    &     \textbf{3}    &     \emph{7}    &     6 \\
	karate & \textbf{3}     &    \textbf{3}   &      \textbf{3}   &     \emph{4}    &     \emph{4} \\
	UKfaculty & \textbf{5}    &    \emph{10}    &     7     &    7   &      \textbf{5}\\
	USairports & 28     &   \emph{57}      &  40   &      NA    &    \textbf{21}\\
	enron & 11    &   \emph{22}    &    20    &     NA   &     \textbf{10}\\

\end{tabular}
\bigskip
\begin{tabular}{r|ccccc}
	\multicolumn{6}{ c }{\textbf{AIC}} \\
	dataset & fast\_greedy  & infomap  &  label\_prop & spinglass  & louvain \\
	\hline
	rfid & 44721.18 & 55234.60 & 56388.23 & \textbf{42864.79} & 44721.18  \\
	karate & 1736.007 & 1736.007 & 1736.007 & 1711.981 & \textbf{1707.768} \\
	UKfaculty & 23456.25 & \textbf{22464.35} & 23424.31 &  22987.02 & 23456.25\\
	USairports & 1212420 &  1213276 &  1215650    &    NA &  \textbf{1210517}\\
	enron & \textbf{326968.3} & 336849.0 & 373913.1   &     NA & 328924.2\\
	
\end{tabular}
\bigskip
\begin{tabular}{r|ccccc}
	\multicolumn{6}{ c }{\textbf{BIC}} \\
	dataset & fast\_greedy  & infomap  &  label\_prop & spinglass  & louvain \\
	\hline
	rfid & 68161.88 & 78583.04 & 79703.13 & \textbf{66364.19} & 68161.88 \\
	karate & 3684.415 & 3684.415 & 3684.415 & 3674.159 & \textbf{3669.947} \\
	UKfaculty & 63875.97 & \textbf{63133.03} & 63924.94 & 63487.66 & 63875.97\\
	USairports & 5812143 &  5823055  & 5818711  &   NA &  \textbf{5808828} \\
	enron & \textbf{657336.3} & 669038.3 & 705683.6  &      NA & 659185.1 \\

\end{tabular}
\caption{\small 
	Results of the fitting of BCCMs to five real-world graphs, with vertex blocks given obtained from five different community detection algorithms.
	The first table reports information about the 5 different graphs used.
	The second table reports the number of clusters detected by each algorithm for each dataset.
	The algorithm detecting the smallest number of clusters is highlighted in bold, and the algorithm detecting the largest number of clusters is highlighted in italic.
	The third table reports the AICs of the different models computed using the different vertex blocks.
	The fourth table reports the BICs of the different models computed using the different vertex blocks.
	The best model, i.e., the one with the lowest AIC/BIC score respectively is highlighted in bold.
	Because the spinglass algorithm is not suitable for disconnected graphs, no result is reported for this method for the last two real-world graphs.
}\label{tab:emp}
\end{table}

\section{Conclusion}\label{sec:conclusion}
In this article we have presented a novel generative model for clustered graphs: the block-constrained configuration model.
It generalises the standard configuration model of random graphs by constraining edges within blocks, preserving degree distributions.
The BCCM builds on the generalised hypergeometric ensemble of random graph, by giving the propensity matrix $\mathbf\Omega$ a block structure.
The framework provided by gHypEG allows for fast estimation of the parameters of the model.
Moreover, thanks to the fact that the closed form of the probability distribution underlying gHypEG is known, it allows for the generation of random realisations, as well as to the effortless computation of likelihoods, and hence various kind of information criteria and goodness-of-fit measures, such as AIC and BIC.

There are many advantages of the formulation highlighted above.
Firstly, the proposed model seamlessly applies to directed and undirected graphs with or without self-loops.
Moreover, closed-form expressions for the probability distribution defining the model allow for its fast estimation over large graphs.
Finally, model selection, facilitated by the gHypE framework, provides a natural method to quantify the optimal number of blocks needed to model given real-world graph.
The statistical significance of a block structure can be studied performing likelihood-ratio tests~\cite{Casiraghi2016}, or comparing information criteria such as AIC, BIC, or the description length of the estimated models.
Furthermore, within the framework of generalised hypergeometric ensembles block-constrained configuration models can be extended including heterogenous properties of vertices or edges (see~\cite{Casiraghi2017}).

BCCMs open new routes to develop community detection algorithms suitable for applications where degree correction is particularly important, because the effects of degrees are naturally accounted for in the general formulation of generalised hypergeometric ensembles of random graphs.

\paragraph{Acknowledgements.}
The author thanks Frank Schweitzer for his support and valuable comments, and Laurence Brandenberger, Giacomo Vaccario and Vahan Nanumyan for useful discussions.

\end{document}